\documentclass[18 pt,preprint]{elsarticle}
\usepackage{amsmath}
\usepackage{amsthm}
\usepackage{amsfonts}
\usepackage{lineno,hyperref}
\usepackage{mathrsfs}
\usepackage{graphicx}
\usepackage{adjustbox}
\usepackage{xcolor,colortbl}
\usepackage{float}
\usepackage{subfig}
\restylefloat{table}
\definecolor{Gray}{gray}{0.85}
\newtheorem{theorem}{Theorem}
\newtheorem{mydef}{Definition}
\newtheorem{exm}{Example}
\modulolinenumbers[5]
\begin{document}

\title{ A Numerical Method for Pricing Discrete Double Barrier Option by Legendre Multiwavelet }

\author[label1]{Amirhossein Sobhani\corref{cor1}}
\ead{a\_sobhani@mathdep.iust.ac.ir, a\_sobhani@aut.ac.ir}

\address[label1]{School of Mathematics, Iran University of Science and Technology, 16844 Tehran,Iran.}
\cortext[cor1]{Corresponding author}
\author[label2]{Mariyan Milev}
\address[label2]{UFT-PLOVDIV, Department of Mathematics and Physics}
\ead{marianmilev2002@gmail.com}

\begin{frontmatter}

\begin{abstract}
In this Article, a fast numerical numerical algorithm for pricing discrete double barrier option is presented. According to Black-Scholes model, the price of option in each monitoring date can be evaluated by a recursive formula upon the heat equation solution. These recursive solutions are approximated by using Legendre multiwavelets as orthonormal basis functions and expressed in operational matrix form. The most important feature of this method is that its CPU time is nearly invariant when monitoring dates increase. Besides, the rate of convergence of presented algorithm was obtained. The numerical results verify the validity and efficiency of the numerical method. 
\end{abstract}

\end{frontmatter}

\linenumbers
\section{Introduction}
Barrier options play a key role in the price risk management of financial markets.
There are two types of barrier options: single and double. In single case we have one barrier but in double case there are two barriers. A barrier option is called knock-out (knock-in) if it is deactivated (activated) when the stock price touches one of the barriers. If the hitting of barriers by the stock price is checked in fixed dates, for example weakly or monthly, the barrier option is called discrete.
\par
Option pricing as one of the most interesting topics in the mathematical finance has been investigated vastly in the literature. 
Kamrad and Ritchken \cite{kamrad1991multinomial}, Boyle and Lau \cite{boyle1994bumping}, Kwok \cite{kwokmathematical}, Heyen and Kat \cite{heynen1997barrier}, Tian \cite{tian1999pricing} and Dai and Lyuu \cite{dai2010bino} used standard lattice techniques, the binomial and trinomial trees, for pricing barrier options. Ahn et al. \cite{ahn1999pricing} introduce the adaptive mesh model (AMM) that increases the efficiency of trinomial lattices. The Monte Carlo simulation methods were implemented in \cite{andersen1996exact,beaglehole1997going,baldi1999pricing, bertoldi2003monte,kuan2003pricing}.  In \cite{andricopoulos2003universal,milev2010numerical}, numerical algorithms based on quadrature methods were proposed.
\par

 Actually a great variety of semi-analytical methods to price barrier options have been recently developed which are based on integral transforms \cite{fusai2006,broadie2005double,fang2009pricing}, or on the transition probability density function of the process used to describe the underlying asset price \cite{andricopoulos2003universal,milev2010numerical,
 golbabai2014highly,broadie1997continuity,dorfleitner2008pricing,
 fusai2007analysis,skaug2007fast,sullivan2000pricing}.  
 These techniques are very high performing for pricing discretely monitored one and double barrier options and our computational results are in very good agreement with them. We would like to make the following essential remarks. An analytical solution for single barrier option is driven by Fusai et. al. in \cite{fusai2006} where the problem of one barrier is reduced to a Wiener-Hopf integral equation and a given z-transform solution of it. To derive a formula for continuous double barrier knock-out and knock-in options Pelsser inverts analytically the Laplace transform by a contour integration \cite{pelsser2000pricing}. Broadie et. al. have found an explicit correction formula for discretely monitored option with one barrier \cite{broadie1997continuity}. However, these three well-known methods \cite{fusai2006,broadie1997continuity, pelsser2000pricing} have not been still applied in the presence of two barriers, i.e. a discrete double barrier option.  
Farnoosh et al. \cite{farnoosh2015,farnoosh2015a} presented numerical algorithms for pricing discrete single and double barrier options with time-dependent parameters. Also, in my last work \cite{amir1} a numerical method for pricing discrete single and double barrier options by projection methods have been presented.

 \par

 This article is organized as follows. In Section 2, the process of finding price of discrete double barrier option under the Black-Scholes model by a recursive formula has bean explained.
 Definition and some features of Legendre multi-wavelets are given in section 3. In section 4, Legendre multi-wavelet expansion is implemented for pricing of discrete double barrier option.
 Finally, numerical results are given in section 5 to confirm efficiency of proposed method.

\section{The Pricing Model}
We assume that the stock price process follows geometric Brownian motion:
	\[d{S_t} = \hat{r}{S_t} dt + \sigma {S_t}d{B_t}\]
where \({S_0}\), \(\hat{r}\) and \(\sigma\) are  initial stock price, risk-free rate and volatility respectively. 
 We consider the problem of pricing knock-out discrete double barrier call option, i.e. a call option that becomes worthless if the stock price touches either lower or upper barrier at the predetermined monitoring dates: \[0 = {t_0} < {t_1} <  \cdots  < {t_M} = T.\]
  If the barriers are not touched in monitoring dates, the pay off at maturity time is $max(S_T-E,0)$, where $E$ is exercise price. The price of option is defined discounted expectation of pay off at the maturity time. %According to the well-known Black-Scholes framework,
  
 Based on Black-Scholes framework, the option price \(\mathcal{P}\left( {S,t,m - 1} \right)\) as a function of stock price at time \(t \in \left( {{t_{m - 1}},{t_m}} \right)\), satisfies in the following partial differential equations\citep{bjork2009arbitrage} 
  \begin{equation} \label{pde:1}
 - \frac{{\partial \mathcal{P}}}{{\partial t}} + \hat{r} S\frac{{\partial \mathcal{P}}}{{\partial S}} + \frac{1}{2}{\sigma ^2}{S^2}\frac{{{\partial ^2}\mathcal{P}}}{{\partial {S^2}}} - \hat{r}\mathcal{P} = 0,
\end{equation} 
subject to the initial conditions:
  \[ \mathcal{P}\left( {S,{t_0},0} \right) = \left( {S - E} \right){\textbf{1}_{\left(\max (E,L) \le S \le U\right)}}\,\]
	\[ \mathcal{P}\left( {S,{t_m},0} \right) = \mathcal{P}\left( {S,{t_m},m - 1} \right){\textbf{1}_{\left(L \le S \le U\right)}};\,\,m = 1,2,...,M - 1~,\]
where \(\mathcal{P}\left( {S,{t_m},m - 1} \right): = \mathop {\lim }\limits_{t \to {t_m}} \mathcal{P}\left( {S,t,m - 1} \right)\).
	By change of variable  \(z = \ln \left( {\frac{S}{L}} \right) \) the partial differential equation \ref{pde:1} and its initial condition is reduced as follows:
  \begin{equation}
	\label{C_equation}
	-{C_t} + \mu {C_z} + \frac{{{\sigma ^2}}}{2}{C_{zz}} = \hat{r}C
	\end{equation}
	\[C\left( {z,{t_0},0} \right) = L\left( {{e^z} - {e^{{E^*}}}} \right){\textbf{1}_{\left( \delta  \le z \le \theta \right)}}\]
	\[C\left( {z,{t_m},m} \right) = C\left( {z,{t_m},m - 1} \right){\textbf{1}_{\left( 0 \le z \le \theta \right)}}\,;\,\,m = 1,2,...,M - 1\,\,\]
	
	where \(C(z,t,m): = \mathcal{P}(S,t,m);\,{E^*} = \ln \left( {\frac{E}{L}} \right);\,\mu  = \hat{r} - \frac{{{\sigma ^2}}}{2};\,\theta  = \ln \left( {\frac{U}{L}} \right)\) and \( \delta  = \max \left\{ {{E^*},0} \right\}\).
Next, by considering $C\left( {z,{t_m},m} \right) = {e^{\alpha z + \beta t}}h(z,t,m)$ where:
	\[\alpha  =  - \frac{\mu }{{{\sigma ^2}}};\,\,{c^2} =  - \frac{{{\sigma ^2}}}{2};\,\beta  = \alpha \mu  + {\alpha ^2}\frac{{{\sigma ^2}}}{2} - \hat{r}.\]
the equation \ref{C_equation} is reduced to the well known heat equation:
	\[ - {h_t} + {c^2}{h_{zz}} = 0\]
	\[h\left( {z,{t_0},0} \right) = L{e^{ - \alpha z}}\left( {{e^z} - {e^{{E^*}}}} \right){\textbf{1}_{\left(\delta  \le z \le \theta \right)}};\,m = 0\]
	\[h\left( {z,{t_m},m} \right) = h\left( {z,{t_m},m - 1} \right){\textbf{1}_{\left( 0 \le \theta  \le z \right)}};\,m = 1,...,M - 1\]
that could be resolved analytically, see e.g \cite{strauss1992partial} as follows;
	\[h(z,t,m) = \left\{ \begin{array}{l}L\int_\delta ^\theta  {{k}\left( {z - \xi ,t} \right){e^{ - \alpha \xi }}\left( {{e^\xi } - {e^{{E^*}}}} \right)d\xi \,;\,\,m = 0} \\\int_0^\theta  {k\left( {z - \xi ,t - {t_m}} \right)h\left( {\xi ,{t_m},m - 1} \right)d\xi } \,;\,\,m = 1,2,...,M-1\end{array} \right.\]
where
\begin{equation}
\label{k1}
\overline{k}(z,t) = \frac{1}{{\sqrt {4\pi {c^2}t} }}{e^{ - \frac{{{z^2}}}{{4{c^2}t}}}}.
\end{equation}
By assuming that monitoring dates are  equally spaced, i.e; \({t_m} = m\tau \) where $\tau =\frac{T}{M}$, \(h\left( {z,{t_m},m - 1} \right)\) is a function of two variables $z$, $m$. Therefore, by defining \({\overline{f}_m}\left( z \right) := h(z,{t_m},m - 1)\), we have:	
\begin{equation}
\label{f1b}
{\overline{f}_1}(z) = \int_0^\theta  {\overline{k}(z - \xi ,\tau){\overline{f}_0}\left( \xi  \right)d\xi } 
\end{equation}
\begin{equation}
\label{fnb}
{\overline{f}_m}(z) = \int_0^\theta  {\overline{k}(z - \xi ,\tau){\overline{f}_{m - 1}}\left( \xi  \right)d\xi } ;\,m = 2,3,...,M
\end{equation}
	
where
\begin{equation}
\label{f0b}
\overline{f}_0\left( z  \right) = L{e^{ - \alpha z}}\left( {{e^z} - {e^{{E^*}}}} \right){\textbf{1}_{\left( \delta  \le z  \le \theta \right)}}.
\end{equation}
%Since we want to use Legendre multiwavelete on interval $[0,1]$ 
by defining
$f_m(z):= \overline{f}_m( \theta z)$ and
\begin{equation}
\label{k}
k(z,\tau):=\theta \overline{k}(\theta z,\tau)=\frac{1}{{\sqrt {4\pi {c^2}t} }}{e^{ - \frac{{{(\theta z)^2}}}{{4{c^2}t}}}}
\end{equation}
 we reach the following relations from \ref{f1b},\ref{fnb} and \ref{f0b}:
\begin{equation}
\label{f1}
{f}_1(z) = \int_0^1  {k(z - \xi ,\tau){f_0}\left( \xi  \right)d\xi } 
\end{equation}
\begin{equation}
\label{fn}
{f_m}(z) = \int_0^1  {k(z - \xi ,\tau){f_{m - 1}}\left( \xi  \right)d\xi } ;\,m = 2,3,...,M
\end{equation}
where
\begin{equation}
\label{f0}
f_0\left( z  \right) = L{e^{ - \alpha \theta z}}\left( {{e^{\theta z}} - {e^{{E^*}}}} \right){\textbf{1}_{\left( \frac{\delta}{\theta} \le z  \le 1 \right)}}.
\end{equation}
which helps us to use Legendre multiwavelete on interval $[0,1]$.
\section{Legendre Multiwavelet}

Let $ L^{2}([0,1]) $ be the Hilbert space of all square-integrable functions on interval $[0,1]$ with the inner product
 \[<f,g>:=\int_{0}^{1}f(x)g(x)dx\]
 and  the norm $\Vert f \Vert=\sqrt{<f,f>}$. An orthonormal multi resolution analysis (MRA) with multiplicity $r$ of $ L^{2}([0,1]) $ is defined as follows:
\begin{mydef} A chain of closed functional subspaces $V_j, j\geq 0$ of $L^{2}([0,1])$ is called orthonormal multi resolution analysis of multiplicity $r$ if:
\begin{enumerate}[(i)]
\item $V_{j}\subset V_{j+1},j \geq 0 $. 
\item $\bigcup\limits_{ j \geq 0}V_{j}$ is dense in $ L^{2}([0,1]) $,i.e.$\overline{\bigcup\limits_{ j \geq 0}V_{j}}=L^{2}([0,1])$.
%\item $f(x) \in V_{j} \Longleftrightarrow f(2x) \in V_{j+1}$.
%\item $f(x) \in V_{j}\Longleftrightarrow f(x-2^{-j}k) \in V_{j}$ for $k=0,...,2^{j}-1$.
\item \label{333} There exists a vector of orthonormal functions $\Phi=[\phi^{0},...,\phi^{r-1}]^{T} $ in $L^{2}([0,1])$, that is called multiscale vector, such that
$\lbrace \phi^{l}_{j,k}:=2^{j/2}\phi^{l}(2^{j}x-k); 0 \leq l \leq r-1, 0 \leq k \leq 2^{j}-1\rbrace$ form an orthonormal basis for $V_{j}$.
\end{enumerate}
\end{mydef}
Now let wavelet space $W_{j}$ be subspace of $V_{j+1}$ such that $V_{j+1}=V{j}\oplus W_{j}$ and $V{j} \perp W_{j}$, i.e. the orthogonal complement of $V_{j}$ in $V_{j+1}$, so we have
\begin{equation}
\label{vwbasisj}
V_{j}=V_{0}\oplus W_{0} \oplus W_{1}\oplus ...W_{j-1}
\end{equation}
\begin{equation}
\label{vwbasisL}
L^{2}([0,1])=V0 \oplus \bigoplus \limits_{j=0}^{\infty}W_{j}.
\end{equation}
The property\ref{333} of MRA shows that $dim(V_{j})=dim(W_{j})=r2^{j}$. Let the function vector $\Psi=[\psi^{0},...,\psi^{r-1}]$ be vector of orthonormal basis of $W_{0}$, that is called multiwavelet vector, then the structure of MRA implies that 
\begin{equation}
W_{j}=\overline{span\lbrace \psi^{l}_{j,k}; 0 \leq l \leq r-1, 0 \leq k \leq 2^{j}-1\rbrace},
\end{equation}
where $\psi^{l}_{j,k}:=2^{j/2}\psi^{l}(2^{j}x-k)$. According to \ref{333} and \ref{vwbasisj} for any $V_{j}$ we have two orthonormal basis set as follow:
\begin{equation}
\Phi_{j}(x)=[\phi_{j,0}^{0}(x),...,\phi_{j,0}^{r-1}(x),...,\phi_{j,2^j-1}^{0}(x),...,\phi_{j,2^j-1}^{r-1}(x)]
\end{equation}
\begin{multline}
\Psi_{j}(x)=[\phi_{0,0}^{0}(x),...,\phi_{0,0}^{r-1}(x),\psi_{0,0}^{0}(x),...,\psi_{0,0}^{r-1}(x),...,\\ \psi_{j-1,0}^{0}(x),...,\psi_{j-1,0}^{r-1}(x),...,\psi_{j-1,2^{j-1}-1}^{0}(x),...,\psi_{j-1,2^{j-1}-1}^{r-1}(x)]
\end{multline}
 From relation \ref{vwbasisL} for any $f \in L^2([0,1])$ we have
\begin{equation}
\label{expanwavelet}
f(x)=\sum_{l=0}^{r-1}c_l \phi^{l}(x)+\sum_{j=0}^{\infty}\sum_{k=0}^{2^{j-1}}\sum_{l=0}^{r-1} c_{j,k} \psi_{j,k}^{l}(x)
\end{equation}
where $c_l=\int_{0}^{1}f(x)\phi^l(x)dx$ and $c_{j,k}=\int_{0}^{1}f(x)\psi_{j,k}^l(x)dx$.\\
Now we define orthonormal projection operator $P_J:L^2([0,1])\rightarrow V_{J}$ as follows:
\begin{equation}
\label{orthoexpan}
P_J(f):=\sum_{l=0}^{r-1}c_l \phi^{l}(x)+\sum_{j=0}^{J-1}\sum_{k=0}^{2^{j-1}}\sum_{l=0}^{r-1} c_{j,k} \psi_{j,k}^{l}(x)
\end{equation}
or equivalently
\begin{equation}
P_J(f):=\sum_{k=0}^{2^{J}}\sum_{l=0}^{r-1} d_{J,k} \phi_{J,k}^{l}(x)
\end{equation}
where $d_{j,k}=\int_{0}^{1}f(x)\phi_{j,k}^{l}(x)dx$. In order to simplify notation, we denote the i-th element of $\Psi_j(x)$ by $\overline{\psi}_i(x)$, so: 
\begin{equation}
\Psi_j(x)=[\overline{\psi}_1(x),\overline{\psi}_2(x),...,\overline{\psi}_{2^j}(x)]
\end{equation}
and then we can rewrite \ref{orthoexpan}:
\begin{equation}
P_J(f):=\sum_{i=0}^{2^J} a_i \overline{\psi}_{i}(x)=\Psi_j(x)'F
\end{equation}
where $a_i=\int_{0}^{1}f(x)\overline{\psi}_i(x)dx$ and $F=[a_1,...,a_{2^j}]$.
 From relation \ref{expanwavelet} $P_J$ is convergence pointwise to identity operator $I$, i.e. 
\begin{equation}
\forall f\in L^{2}[0,\theta]~~~\mathop {\lim }\limits_{J \to \infty } \left\| {{P_J}(f) - f} \right\| = 0.
\end{equation}

We use Legendre polynomial to construct Legendre Multiwavelet that has introduced by Alpert in \cite{alpert1993class}.
Legendre polynomial, \({p_i}(x)\), is defined as follows
	\[{p_0}(x) = 1\,\,\,\,,\,\,\,{p_1}(x) = x\]
with the following recurrence formula:
	\[{p_i}(x) = x{p_{i - 1}}(x) + \left( {\frac{i}{{i + 1}}} \right)\left( {x{p_{i - 1}}(x) - {p_{i - 2}}(x)} \right)\]
The \(\left\{ {{p_i}(x)} \right\}_{i = 0}^\infty \) is an orthogonal basis for \({L^2}([ - 1,1])\).\\
We define $V_{j}$ as follows
\begin{equation}
V_{j}:=\lbrace f \vert ~f~ be~ a~ polynomial~ of~ degree \leq r~ on~ I_{i}, 1 \leq i \leq 2^{j} \rbrace
\end{equation}
where $I_{i}:=[2^{-j}(i-1),2^{-j}i)$. It is obvious that $V_{j}\subset V_{j+1}$ and $\overline{\bigcup\limits_{ j \geq 0}V_{j}}=L^{2}([0,1])$. Now let $\phi^{l}$ be a Legendre multiscaling function, that is defined as
\begin{equation}
\phi^{l}:=
 \left\{ \begin{array}{ll}
\sqrt{2l+1}p_{l}(2x-1)& x \in [0,1),\\
0,     & o.w,
\end{array} \right. 
\end{equation}
and $\Phi:=[\phi^{0},...,\phi^{r-1}]^{T} $ be the multiscale vector. It is easy to verify that
\begin{equation}
\lbrace \phi^{l}_{j,k}:=2^{j/2}\phi^{l}(2^{j}x-k); 0 \leq l \leq r-1, 0 \leq k \leq 2^{j}-1\rbrace,
\end{equation}
forms an orthonormal basis for $V_j$. Now let $\Psi=[\psi^{1},...,\psi^{r-1}]$ be the Legendre multiwavelet vector. Because of $W_0 \subset V_1$ each $\psi_{l}$ could be expanded as follows:
\begin{equation}
\label{twoscale}
\psi^{l}= \sum_{k=0}^{r-1} g_{l,k}^{0} \phi^{k}(2x)+ \sum_{k=0}^{r-1} g_{l,k}^{1} \phi^{k}(2x-1)~~, ~~ 0 \leq l \leq r-1
\end{equation}
In addition, $W_0 \perp V_0$ and $1,x,..,x^{r-1} \in V_0$, so the first $r$ moment of $\lbrace\psi^{l} \rbrace_{l=0}^{r-1}$ vanish:
\begin{equation}
\label{vanishing}
 \int_{0}^{1} \psi^{l}(x)x^{i}dx=0~~~  0\leq l,i \leq r-1
 \end{equation} 
 on the other hand, we have
 \begin{equation}
 \label{psiortho}
 \int_{0}^{1} \psi^{i}(x)\psi^{j}(x)dx=0~~~  0\leq i,j \leq r-1
 \end{equation}
 so for finding $2r^2$ unknown coefficients $g_{i,j}$ in \ref{twoscale}, it is enough to solve $2r^2$ equations \ref{vanishing} and \ref{psiortho}. If $f \in L^2([0,1])$ be $k$ times differentiable, the following theorem about bound of error is obtained \cite{alpert1993class}:
\begin{theorem}
Suppose that the real function $f \in C^{r}([0,1])$. Then $P_J(f)$ approximates $f$ with the following error bound:
\begin{equation}
\label{rate1d}
  \Vert P_J(f)-f \Vert \leq \frac{2^{(-Jr+1)}}{4^{r}r!} \sup_{x\in [0,1]} \vert f^{r}(x) \vert.
\end{equation}  
\end{theorem}  
Legendre multiscaling and multiwavelet functions are presented for $r=4$ as follows \cite{lakestani2011numerical}:
\begin{equation}
\begin {array}{l r }
\phi^0(x)=   1&0\leq x < 1\\
\phi^1(x)= \sqrt {3} \left( 2\,x-1 \right) &0 \leq x < 1\\
\phi^2(x)= \sqrt {5} \left( 6\,{x}^{2}-6\,x+1 \right)  &0 \leq x < 1 \\
\phi^3(x)= \sqrt {7} \left( 20\,{x}^{3}-30\,{x}^{2}+12\,x-1 \right)&0 \leq  x < 1
\end {array}
\end{equation}
\begin{equation}
\begin {array}{l  }
\psi^0(x)=\begin{cases} -\sqrt {\frac{15}{34}} \left( 224\,{x}^{3}-216\,{x}^{2}+56\,x-3 \right)   &0 \leq x< 1/2 \\ \sqrt {\frac{15}{34}}  \left( 224\,{x}^{3}-456\,{x}^{2}+296\,x-61 \right)  & 1/2 \leq x < 1/2 \end{cases}\\ \noalign{\medskip}
\psi^1(x)=\begin{cases} -\sqrt {\frac{1}{21}} \left( 1680\,{x}^{3}-1320\,{x}^{2}+270\,x-11 \right) &0 \leq x\le 1/2 \\   \sqrt {\frac{1}{21}}\left( 1680\,{x}^{3}-3720\,{x}^{2}+2670\,x-619 \right) & 1/2 \leq x < 1/2 \end{cases}\\ \noalign{\medskip}
\psi^2(x)=\begin{cases} -\sqrt {\frac{35}{17}} \left( 256\,{x}^{3}-174\,{x}^{2}+30\,x-1 \right) &0 \leq x< 1/2 \\  \sqrt {\frac{35}{17}} \left( 256\,{x}^{3}-594\,{x}^{2}+450\,x-111 \right) & 1/2 \leq x< 1/2 \end{cases}\\ \noalign{\medskip}
\psi^3(x)=\begin{cases} \sqrt {\frac{5}{42}}\left( 420\,{x}^{3}-246\,{x}^{2}+36\,x-1 \right) &0 \leq x< 1/2 \\  \sqrt {\frac{5}{42}} \left( 420\,{x}^{3}-1014\,{x}^{2}+804\,x-209 \right) & 1/2 \leq x < 1/2 \end{cases}

\end {array}
\end{equation}
\section{Pricing by Legendre Multiwavelet}
Let operator $\mathcal{K}:L^{2}([0,1]) \to L^{2}([0,1 ])$ is defined as follows:
\begin{equation}
\label{operator}
\mathcal{K}\left( f \right)(z): = \int_0^1  {\kappa (z - \xi ,\tau )f(\xi )} d\xi .
\end{equation}
where $\kappa$ is defined in \ref{k}. Because \(\kappa \) is a continuous function, \(\mathcal{K} \) is a bounded linear compact operator on \(L^{2}([0,1])\)\cite{atkinson2005theoretical,krasnosel1984geometrical}. According to the definition of operator $\mathcal{K} $, equations \ref{f1} and \ref{fn} can be rewritten as below:
\begin{equation}
f_1=\mathcal{K}f_{0}
\end{equation}
\begin{equation}
f_m=\mathcal{K}f_{m-1}~~\,m = 2,3,...,M
\end{equation}
We denote
\begin{equation}
{{\tilde f}_{1,J}} = {P_J}\mathcal{K}\left( {{f_0}} \right)
\end{equation}
\begin{equation}
\label{recfmadm}
{{\tilde f}_{m,J}} = {P_J}\mathcal{K}\left( {{{\tilde f}_{m - 1,J}}} \right)= \left({P_J}\mathcal{K}\right)^{m}\left(f_{0}\right)\,,\,m \ge 2.
\end{equation}
where ${P_J}\mathcal{K}$ is as follows:
\[({P_J}\mathcal{K})(f) = {P_J}\left( {\mathcal{K}(f)} \right).\]
Since the continuous projection operators \({P_J}\) converge pointwise to identity operator \(I\), then operator \({P_J}\mathcal{K}\) is also a compact operator and 
\begin{equation}
\label{first-convergency}
\mathop {\lim }\limits_{n \to \infty }\left\| {{P_J}\mathcal{K} - \mathcal{K}} \right\| = 0
\end{equation}
(see \cite{anselone1968spectral}). With attention to the following inequality 
\begin{equation}
\label{37777}
\left\| {\left( {{P_J}\mathcal{K}} \right)}^m-{\mathcal{K}^m} \right\| \leq \Vert \left( {{P_J}\mathcal{K}} \right) \Vert \Vert {\left( {{P_J}\mathcal{K}} \right)}^{m-1}-{\mathcal{K}^{m-1}} \Vert - \Vert {{P_J}\mathcal{K} - \mathcal{K}} \Vert \Vert {\mathcal{K}}\Vert^{m-1}
\end{equation}
and relation \ref{first-convergency} by induction we get
\begin{equation}
\label{second-convergency}
\mathop {\lim }\limits_{n \to \infty }\left\| {\left( {{P_J}\mathcal{K}} \right)}^m-{\mathcal{K}^m} \right\| = 0.
\end{equation}
Therefore, the following convergence result is concluded:
 \begin{equation}
 \label{error}
 \left\| {{{\tilde f}_{m,J}} - {f_{m}}} \right\| =\left\|   {\left( {{P_J}\mathcal{K}} \right)}^m  \left(f_{0} \right)
     -   \mathcal{K}^m     \left(f_{0} \right)        \right\| 
\le \left\| {{{\left( {{P_J}\mathcal{K}} \right)}^m} - {\mathcal{K}^m}} \right\|\left\| {{f_0}} \right\|  \to 0 as\,\,J \to \infty.
\end{equation}
From \ref{37777} and \ref{error}, we infer that the rate of convergence  ${\tilde f}_{m,J}$ to $   {f_{m}}$ and ${{P_J}\mathcal{K}}$ to $\mathcal{K}$ are the same. Using the relation \ref{rate1d} and properties of integral operator $\mathcal{K}$, it is easy to confirm that
 \begin{equation}
 \label{rate2}
 \left\| {{P_J}\mathcal{K} - \mathcal{K}} \right\| \leq \frac{2^{(-Jr+1)}}{4^{r}r!} \sup_{z,\xi\in [0,1]} \vert \frac{\partial \kappa (z - \xi ,\tau )}{\partial z^r} \vert.
 \end{equation}

Since, \({\tilde f_{m,J}} \in {V_J}\) for \( m \geq 1\), we can write 
	\[{\tilde f_{m,J}} = \sum\limits_{i = 0}^{{r2^{J}}} {{a_{mi}}{\overline{\psi} _i}(z)}  = {\Psi '_J}(x)F_{m},\] 	
where \({ F_m} = [{a_{m0}},{a_{m1}}, \cdots ,{a_{m{2^j}}}]'\).
From equation \ref{recfmadm} we obtain
\begin{equation}
\label{19}
{\tilde f_{m,J}} = {({P_J}\mathcal{K})^{m - 1}}{\left( \tilde f_{1,J} \right)}.
\end{equation}
Since \({V_J}\) is a finite dimensional linear space, thus the linear operator \({P_J}\mathcal{K}\) on \({V_J}\) could be considered as a \({r2^J} \times {r2^J}\) matrix \(K\). Consequently equation \ref{19} can be written as following matrix operator form
\begin{equation}
\label{fmatrix}
{\tilde f_{m,J}} = {\Psi '_J}{K^{m - 1}}{ F_1}.
\end{equation}

For evaluation of the option price by \ref{fmatrix}, it is enough to calculate the matrix operator \(K\) and the vector \({ F_1}\). It is easy to check (see \cite{amir1}) that:
\[{F_1} = [{a_{11}},{a_{12}}, \cdots ,{a_{1r2^J}}]' \]
\[K = {\left( {{k_{ij}}} \right)_{{r2^J} \times {r2^J}}}\]
where
\[{a_{1i}} = \int_0^1 {\int_{\delta/\theta } ^1 {{\overline{\psi}_i}(\eta )\kappa (\eta  - \xi ,\tau ){f_0}(\xi )d\xi d\eta } \,\,} ,\,\,0 \le i \le r2^J.\]
\[{k_{ij}} = \int_0^1  {\int_0 ^1  {{\overline{\psi} _i}(\eta ){\overline{\psi} _j}(\xi )\kappa (\eta  - \xi ,\tau )d\xi d\eta } \,\,} .\]

Therefore,the price of the knock-out discrete double barrier option can be estimated as follows:
	\begin{equation}
	\label{fNmatrix}
	\mathcal{P}\left( {S_0,t_M,M - 1}\right) \simeq {e^{\alpha z_0 + \beta t}}{\tilde f_{M,J} \left( z_0/\theta  \right) }
	\end{equation}
where $z_0=\log\left( \frac{S_0}{L}\right)$ and $\tilde f_{M,n}$ from \ref{fmatrix}.
The matrix form of relation \ref{fmatrix} implies that the computational time of presented algorithm be nearly fixed when monitoring dates increase. Actually, if we set $N=r2^J$ the complexity of our algorithm is $\mathcal{O}{(N^2)}$ that dose not depend on number of monitoring dates.

\section{Numerical Result}
\begin{figure}
\subfloat[$r=3$]{
\includegraphics[width=.5\linewidth]{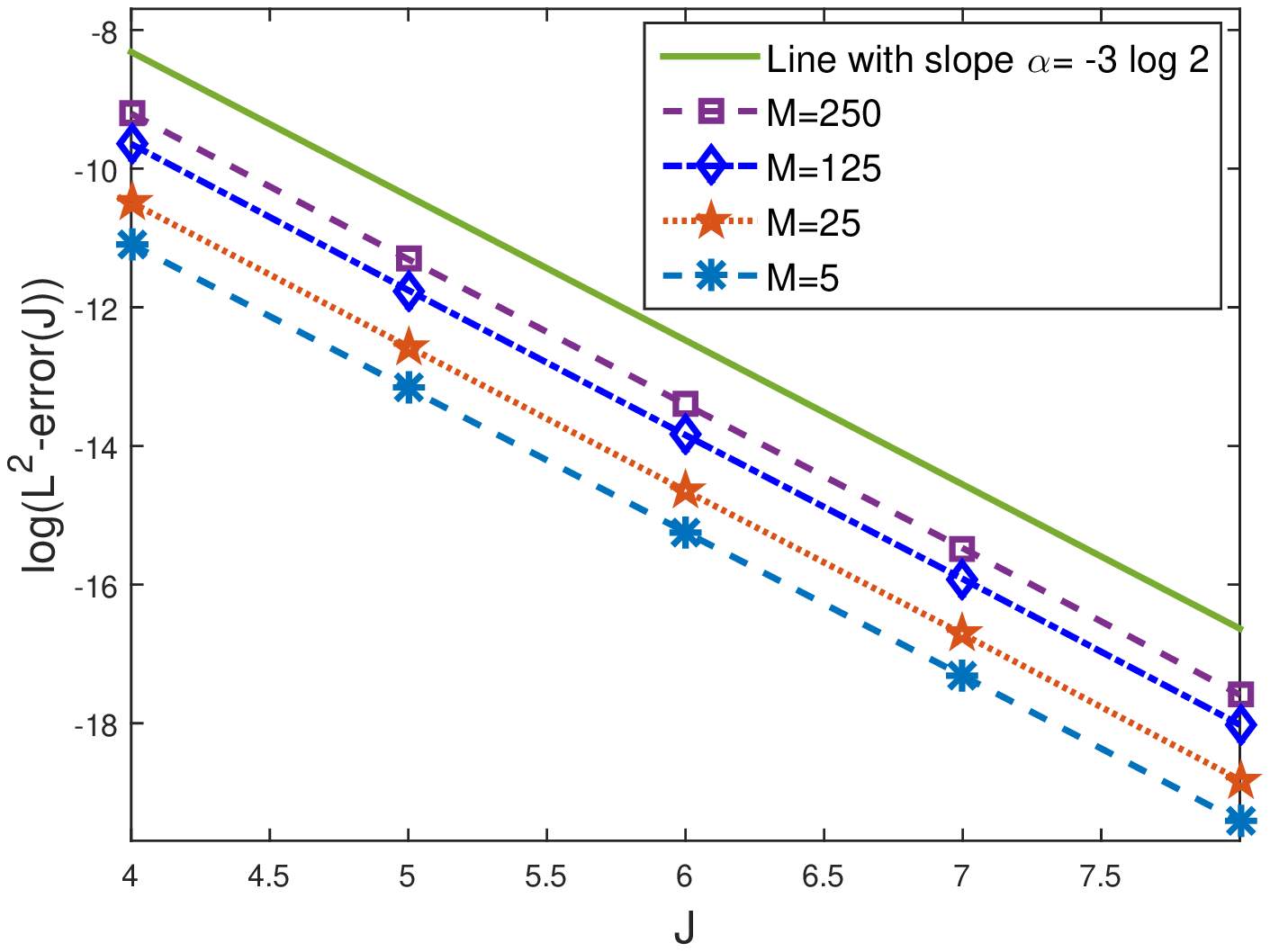}
}\hfill
\subfloat[$r=4$]{
\includegraphics[width=.5\linewidth]{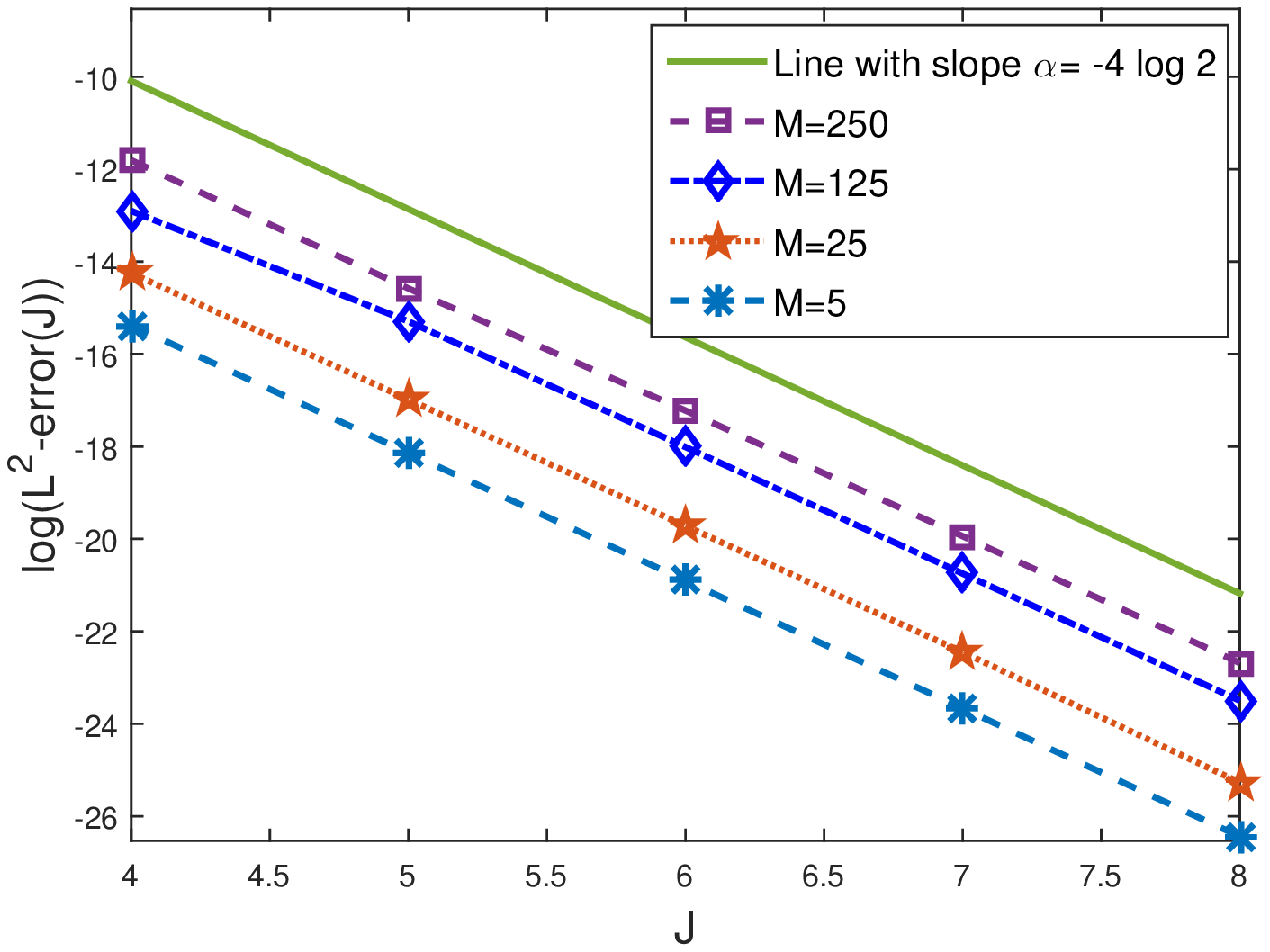}
}
\caption{$log(L^{2}-error(J))$ for example 1 with L=95}
\label{fig:figsdd}
\end{figure}

In the current section, the presented method in previous section for pricing knock-out call discrete double barrier option is compared with some other methods. The numerical results are obtained from the relation \ref{fNmatrix} with $r2^{J}$ basis functions. In the following we denote $\left\| {{{\tilde f}_{m,J}} - {f_{m}}} \right\| $ by ${e_{2}(J)}$ and $L^{2}-error(J)$. As we discussed in the previous section, the rate of convergence  ${\tilde f}_{m,J}$ to $   {f_{m}}$ and ${{P_J}\mathcal{K}}$ to $\mathcal{K}$ are the same. Therefore, ${e_{2}(J-1)}/{e_{2}(J)}$ must be about $2^r$ from \ref{rate2}. In addition, relation \ref{rate2} implies that the slope of $log(L^{2}-error(J))$ be about $\alpha= -r log(2)$. Source code has been written in \textsc{Matlab} 2015 on a 3.2 GHz Intel Core i5 PC with 8 GB RAM.  
 \begin{exm}
 \label{example1}
 
 In the first example, the pricing of knock-out call discrete double barrier option is considered with the following parameters: $r=0.05$, $\sigma=0.25$, $T=0.5$, $S_0=100$, $E=100$, $U=120$ and $L=80~,90~,95~,99~,99.5$. In table \ref{tab1}, numerical results of presented method with Milev numerical algorithm \citep{milev2010numerical}, Crank-Nicholson \cite{wade2007smoothing}, trinomial, adaptive mesh model (AMM) and quadrature method QUAD-K200 as benchmark \cite{shea2005numerical} are compared for various number of monitoring dates. In addition, it can be seen that CPU time of presented method is fixed against increases of monitoring dates. The $L^{2}-error(J)$ are demonstrated for $L=90$ and $M=250$ in Table \ref{tab2} which results verify the convergence rate of our algorithm. Fig.\ref{fig:figsdd} shows the plot  of $log(L^{2}-error(J))$ for $r=3,4$ and it can be seen that the slope of $log(L^{2}-error(J))$ is near to $\alpha= -r log(2)$.

\end{exm}

\begin{table}[H]
\caption{Double barrier option pricing of Example \ref{example1}: $T = 0.5$, $r= 0.05$, $\sigma = 0.25$, $S_0 = 100$, $E = 100$.}
\label{tab1}
\begin{tabular}{c c c c c c c c } 
\hline 
\multicolumn{1}{c}{M}  & \multicolumn{1}{c}{L} &\begin{tabular}[c]{@{}c@{}}Presented\\ Method\\ (r=4 , J=5)\end{tabular} & \begin{tabular}[c]{@{}c@{}}Milev\\ ($200$)\end{tabular}& \begin{tabular}[c]{@{}c@{}}Milev\\ ($400$)\end{tabular} &\multicolumn{1}{c}{Trinomial}&\multicolumn{1}{c}{AMM-8}& \multicolumn{1}{c}{Benchmark}   \\ 
\hline 
& \multicolumn{5}{c}{} \\  
  & 80   &2.4499 &-	     &-       &2.4439 & 2.4499&2.4499 \\ 
%\cline{2-8} 
  & 90   &2.2028 &-	     &-      &2.2717& 2.2027 &2.2028 \\ 
%\cline{2-8}
5 & 95   &1.6831 &1.6831 &1.6831 &1.6926&1.6830 &1.6831 \\ 
%\cline{2-8} 
  & 99   &1.0811 &1.0811 &1.0811 &0.3153&1.0811 &1.0811 \\ 
%\cline{2-8}
  & 99.9 &0.9432 &0.9432 &0.9432 &-&0.9433 &0.9432 \\ 
%\cline{2-8}
\rowcolor[HTML]{C0C0C0} 
 CPU& \- & 0.25 s &1 s & 5 s &\multicolumn{3}{c}{}  \\ 
%\cline{2-8} 
   & 80   &1.9420 &-      &-     &1.9490&1.9419 &1.9420 \\ 
%\cline{2-8} 
   & 90   &1.5354 &-      &-     &1.5630&1.5353 &1.5354 \\ 
%\cline{2-8}
25 & 95   &0.8668 &0.8668 &0.8668 &0.8823&0.8668&0.8668 \\ 
%\cline{2-8} 
   & 99   &0.2931 &0.2931 &0.2931 &0.3153&0.2932&0.2931 \\ 
%\cline{2-8}
   & 99.9 &0.2023 &0.2023 &0.2023 &-&0.2024&0.2023 \\ 
%\cline{2-8} 
\rowcolor[HTML]{C0C0C0} 
 CPU& \- & 0.25 s& 8 s & 30 s &\multicolumn{3}{c}{}  \\ 
%\cline{2-8} 
 & 80 & 1.6808 &- &- &1.7477&1.6807& 1.6808  \\ 
%\cline{2-8}
 & 90 & 1.2029 &- &- &1.2370&1.2028& 1.2029  \\ 
%\cline{2-8}
125& 95 & 0.5532 &0.5528 &0.5531  &0.5699&0.5531 & 0.5532  \\ 
%\cline{2-8}
& 99 & 0.1042 &0.1042 &0.1042  &0.1201&0.1043& 0.1042  \\ 
%\cline{2-8}
 & 99.9 & 0.0513 &0.0513 &0.0513&-&0.0513 & 0.0513  \\ 
%\cline{2-8}
\rowcolor[HTML]{C0C0C0} 
CPU& \- & 0.25 s& 35 s & 150 s &\multicolumn{3}{c}{}  \\ 
%\cline{2-8}
    & 80   &1.6165 &- &-  &1.8631&1.6163& 1.6165 \\ 
%\cline{2-8}
    & 90   &1.1237 &- &-  &1.2334&1.1236& 1.1237  \\ 
%\cline{2-8}
250 & 95   &0.4867 &- &-  &0.5148&0.4867 & 0.4867  \\ 
%\cline{2-8}
    & 99   &0.0758 &- &-  &0.0772&0.0759& 0.0758  \\ 
%\cline{2-8}
    & 99.9 &0.0311 &- &-  &-&0.0311 & 0.0311  \\ 
%\cline{2-8}
\rowcolor[HTML]{C0C0C0} 
CPU& \- & 0.25 s&\multicolumn{5}{c}{}  \\ 
\end{tabular} 
%}
%\end{adjustbox}
\end{table}

\begin{table}[H]
\centering
\caption{$L^2-error$ of example 1 for $L= 90$ and $M=250$.}
\label{tab2}
\begin{tabular}{c|c|c|c|c|}
\cline{2-5}
\multicolumn{1}{l|}{}   & \multicolumn{2}{c|}{r=3}                    & \multicolumn{2}{c|}{r=4}                     \\ \hline
\multicolumn{1}{|c|}{J} & $e_{2}(J)$  & ${e_{2}(J-1)}/{e_{2}(J)}$ & $e_{2}(J)$   & ${e_{2}(J-1)}/{e_{2}(J)}$ \\ \hline
\multicolumn{1}{|c|}{4} & 1.00241 e-4 & -                             & 7.45781 e -6 & -                             \\ \hline
\multicolumn{1}{|c|}{5} & 1.22740 e-5 & 8.16                          & 4.65569 e-7  & 16.01                         \\ \hline
\multicolumn{1}{|c|}{6} & 1.50805 e-6 & 8.14                          & 3.31567 e-8  & 14.04                         \\ \hline
\multicolumn{1}{|c|}{7} & 1.90330 e-7 & 7.92                          & 2.18567 e-9  & 15.16                         \\ \hline
\multicolumn{1}{|c|}{8} & 2.29513 e-8 & 8.29                          & 1.40662 e-10 & 15.53                         \\ \hline
\end{tabular}
\end{table}

\begin{figure}
\subfloat[$r=3$]{
\includegraphics[width=.5\linewidth]{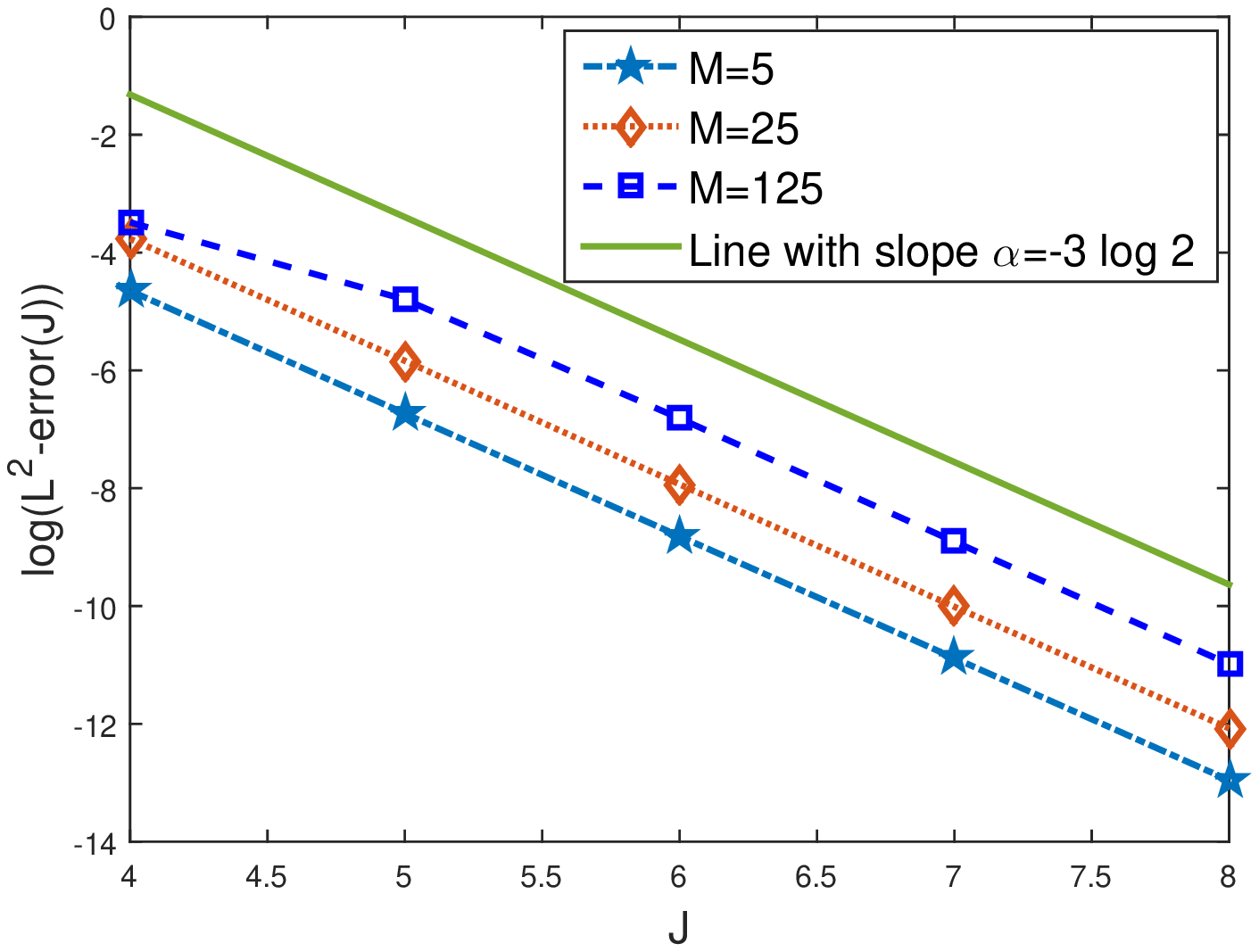}
}\hfill
\subfloat[$r=4$]{
\includegraphics[width=.5\linewidth]{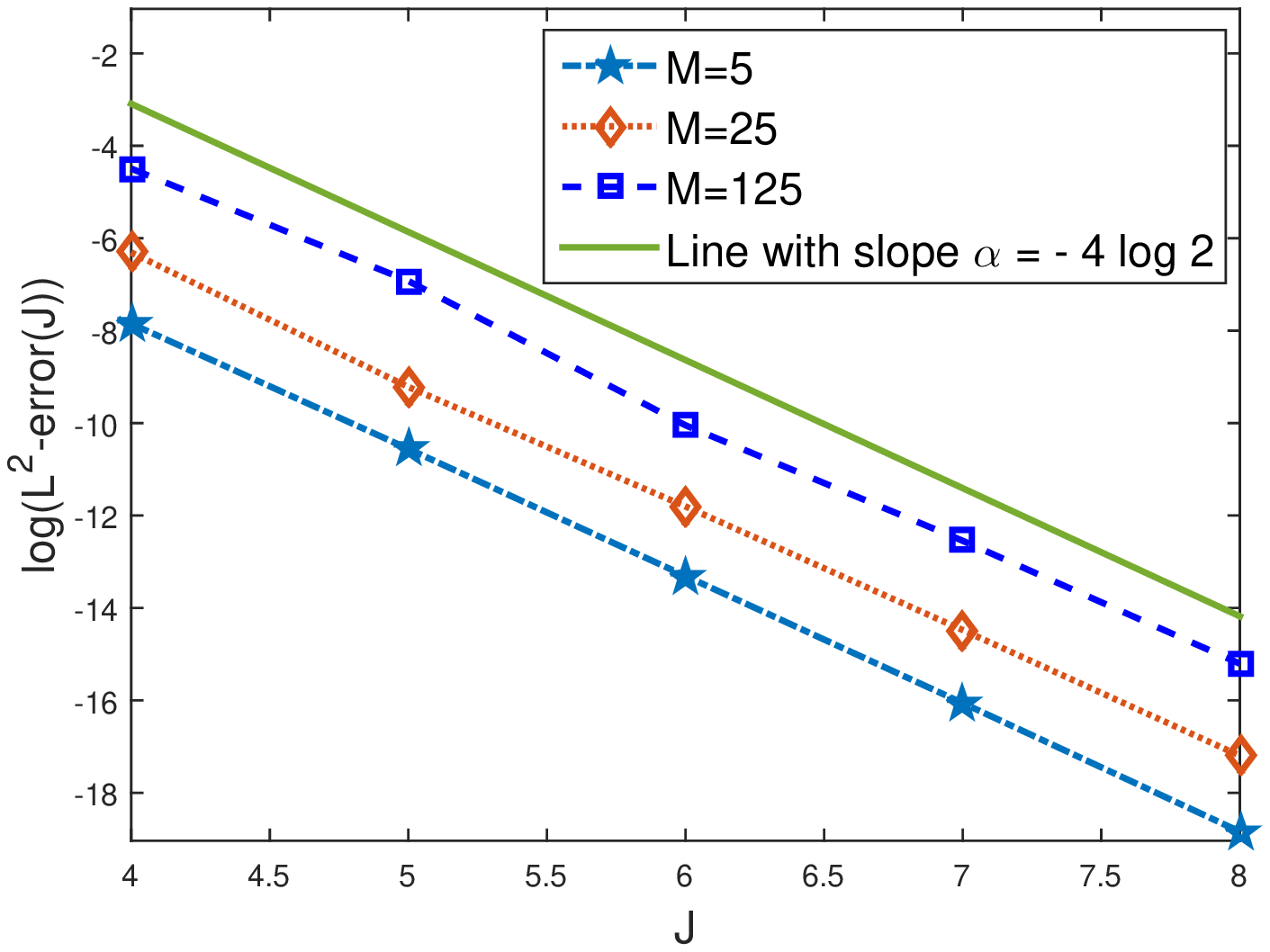}
}
\caption{$log(L^{2}-error(J))$ for example 3 with L=95}
\label{fig:fig2}
\end{figure}

\begin{exm}
\label{example2}
In this example, the parameters of knock-out call discrete double barrier option is considered as $r=0.05$, $\sigma=0.25$, $T=0.5$, $E=100$, $U=110$ and $L=95$. In table \ref{exam2} the option price for different spot prices are evaluated and compared with Milev numerical algorithm \citep{milev2010numerical}, Crank-Nicholson \cite{wade2007smoothing} and the Monte Carlo (MC) method with $10^7$ paths \cite{brandimarte2003numerical}. 
\end{exm} 
\begin{table}[H]

\caption{Double barrier option pricing of Example \ref{example2}: $T = 0.5$, $M=5$, $r= 0.05$, $\sigma = 0.25$, $E = 100$, $U=110$ and $L=95$.}
\label{exam2}
\resizebox{\columnwidth}{!}{%
\begin{tabular}{cccccc}
\hline
$S_0$    & \begin{tabular}[c]{@{}c@{}}Presented  Method\\ (r=4 , J=5)\end{tabular} & \begin{tabular}[c]{@{}c@{}} Crank-Nicolson \\ (1000) \end{tabular} & \begin{tabular}[c]{@{}c@{}}Milev\\ (1000)\end{tabular} & \begin{tabular}[c]{@{}c@{}}Milev\\ (400) \end{tabular}& \begin{tabular}[c]{@{}c@{}} MC (st.error) \\ with $10^7$ paths\end{tabular} \\ \hline
95       & 0.174498 & 0.1656 & 0.174503 & 0.174498& -\\
95.0001  & 0.174499 & $\simeq$ 0.1656 & 0.174501 & 0.174499 &0.17486 (0.00064)\\
95.5     & 0.182428 & 0.1732   & 0.182429& 0.182428 &0.18291 (0.00066)      \\
99.5     & 0.229349 & 0.2181 & 0.229356 & 0.229349 & 0.22923 (0.00073)\\
100      & 0.232508 & 0.2212  & 0.232514  & 0.232508& 0.23263 (0.00036) \\
100.5    & 0.234972 &0.2236   & 0.234978 & 0.234972 & 0.23410 (0.00073)    \\
109.5    & 0.174462 & 0.1658 & 0.174463 & 0.174462& 0.17426 (0.00063)\\
109.9999 & 0.167394 & $\simeq$ 0.1591 & 0.167399  & 0.167394 & 0.16732 (0.00062)   \\
110      & 0.167393 &0.1591 & 0.167398  & 0.167393 & -\\
\rowcolor[HTML]{C0C0C0} 
CPU      & 0.25 s & Minutes  & 1 s & 39 s&
\end{tabular}
}
\end{table}
\begin{exm}
\label{example3}
Due to the fact that the probability of crossing upper barrier during option's life when $U\geq2E$ is too small, the price of discrete single down-and-out call option can be estimated by double ones by setting upper barrier greater than $2E$ (for more details see\citep{milev2010numerical}). Now, we consider a discrete single down-and-out call option with the following parameters: $r=0.1$,  $\sigma=0.2$, $T=0.5$, $S_0=100$, $E=100$ and $L=95~,99.5~,99.9~$. The price is estimated by double ones with $U=2.5 E$. The numerical results are shown in table \ref{exam3} and compared with Fusai\textsc{\char13}s analytical formula \cite{fusai2006}, the Markov chain method (MCh)\cite{duan2003pricing} and the Monte Carlo method (MC) with $10^8$ paths \cite{bertoldi2003monte} that shows the validity of presented method in this case. Fig.\ref{fig:fig2} shows the plot  of $log(L^{2}-error(J))$ for $r=3,4$ and it can be seen that the slope of $log(L^{2}-error(J))$ is near to $\alpha= -r log(2)$. 
\end{exm}
\begin{table}[H]
\centering
\caption{Single barrier option pricing of Example \ref{example3}: $T = 0.5$, $r= 0.1$, $\sigma = 0.2$, $S_0 = 100$, $E = 100$, $U=250$.}
\label{exam3}
\begin{tabular}{ccccccc}
\hline
L    & M   & \begin{tabular}[c]{@{}c@{}}Presented\\ Method\\ (r=4 ,J=6)\end{tabular} & \begin{tabular}[c]{@{}c@{}}Presented\\ Method\\ (r=4 ,J=7)\end{tabular} & \begin{tabular}[c]{@{}c@{}}Fusai Analytical\\ Method (IR17)\end{tabular} &\begin{tabular}[c]{@{}c@{}} MCh\end{tabular}& \begin{tabular}[c]{@{}c@{}} MC (st.error) \\ with $10^8$ paths\end{tabular}\\ \hline
95   & 25  & 6.63155                                                                    & 6.63156 & 6.63156   &6.6307& 6.63204 (0.0009)                                                        \\
99.5 & 25  & 3.35559 & 3.35558 & 3.35558 & 3.3552& 3.35584 (0.00068)                                                           \\
99.9 & 25  & 3.00887 & 3.00887 & 3.00887 &3.0095&3.00918 (0.00064)                                                           \\
95   & 125 & 6.16864                                                                    & 6.16864 & 6.16864 & 6.1678& 6.16879 (0.00088)                                                           \\
99.5 & 125 & 1.96132 & 1.96130 & 1.96130& 1.9617&1.96142 (0.00053) \\
99.9 & 125 & 1.51019 & 1.51021 & 1.51068 & 1.5138 & 1.5105 (0.00046) 
                                                          \\
 \rowcolor[HTML]{C0C0C0} 
CPU  &    &  0.48 s & 0.83 s& &                                                                                                                                                     &                                                                  
\end{tabular}
\end{table}

\section{Conclusion and remarks}
In this article, we used the Legendre multiwavelet for pricing discrete single and double barrier options. In section 4 we obtained a matrix relation \ref{fmatrix} for solving this problem. Numerical results confirm that growth of computational time is negligible when the number of monitoring dates increase. On the other hand, the rate of convergence of presented algorithm has been obtained theoretically and verified numerically .
 
\section*{References}

\bibliography{mybibfile}

\end{document}